\documentclass[twocolumn]{article}

\usepackage{titling} 
\usepackage{authblk} 
\usepackage{graphicx} 
\usepackage{xcolor} 
\usepackage{url}
\usepackage{hyperref}  
\usepackage[margin=1.25in]{geometry} 

\title{\Large \textbf{Pycro-manager: open-source software for integrated microscopy hardware control and image processing}}

\date{} 

\author[1,2,3,4*]{Henry Pinkard}
\author[5]{Nico Stuurman}
\author[2,3]{Laura Waller}

\affil[1]{Computational Biology Graduate Group, University of California, Berkeley, CA 94720, USA}
\affil[2]{Department of Electrical Engineering and Computer Sciences, University of California, Berkeley, CA 94720, USA}
\affil[3]{Berkeley Institute for Data Science}
\affil[4]{University of California San Francisco Bakar Computational Health Sciences Institute}
\affil[5]{Howard Hughes Medical Institute and University of California San Francisco, San Francisco, CA 94158}

\affil[*]{\normalfont{\textbf{Correspondence: hbp [at] berkeley.edu}}}

\begin{document}

\maketitle

\renewcommand{\abstractname}{\vspace{-26pt }} 

\begin{abstract} \bf

\(\mu\)Manager, an open-source microscopy acquisition software, has been an essential tool for many microscopy experiments over the past 15 years, but is not easy to use for experiments in which image acquisition and analysis are closely coupled. This is because \(\mu\)Manager libraries are written in C++ and Java, whereas image processing is increasingly carried out with data science and machine learning tools most easily accessible through the Python programming language. We present Pycro-Manager, a tool that enables rapid development of such experiments, while also providing access to the wealth of existing tools within \(\mu\)Manager through Python.
\end {abstract}

Cutting-edge innovations in biological microscopy are increasingly blurring the line between data acquisition and data analysis. Often, the captured image requires significant image processing to produce the final image, and users rely on pipelines of multiple programs or software libraries to get from the capture stage to the final image. Computational microscopy and machine learning-based methods take this paradigm to an extreme, often producing raw measurements that aren't human-interpretable without post-processing \cite{Christiansen2018, Ounkomol2018a, Weigert2018, Guo2019} Furthermore, new "adaptive" methods rely on image processing during acquisition to actively control various parameters of the microscope \cite{Royer2016, Pinkard2019}. Testing new ideas and applying them for biological discovery is often impeded by the lack of powerful yet sufficiently flexible control software, necessitating the development of bespoke solutions that work only with a specific instrument. 

In situations where the close coupling of data acquisition and image analysis are not required, \(\mu\)Manager \cite{Edelstein2010a, Edelstein2014} has often been the \textit{de facto} solution. This owes to the existence of a hardware abstraction layer and an extensive library of `device adapters', which allow control of hundreds of different types of hardware ranging from cameras to complete microscopes. Community contributions of device adapters, plugins, and scripts provide a treasure trove of hundreds of developer-years of microscopy automation. However, despite the power of these libraries, which are written in C++ and Java, they are often difficult to integrate with the latest developments in computer vision and scientific computing, which are most readily available through the Python programming language \cite{Virtanen2020a}.

\begin{figure*}[htbp]
\centering
\includegraphics[width=\linewidth]{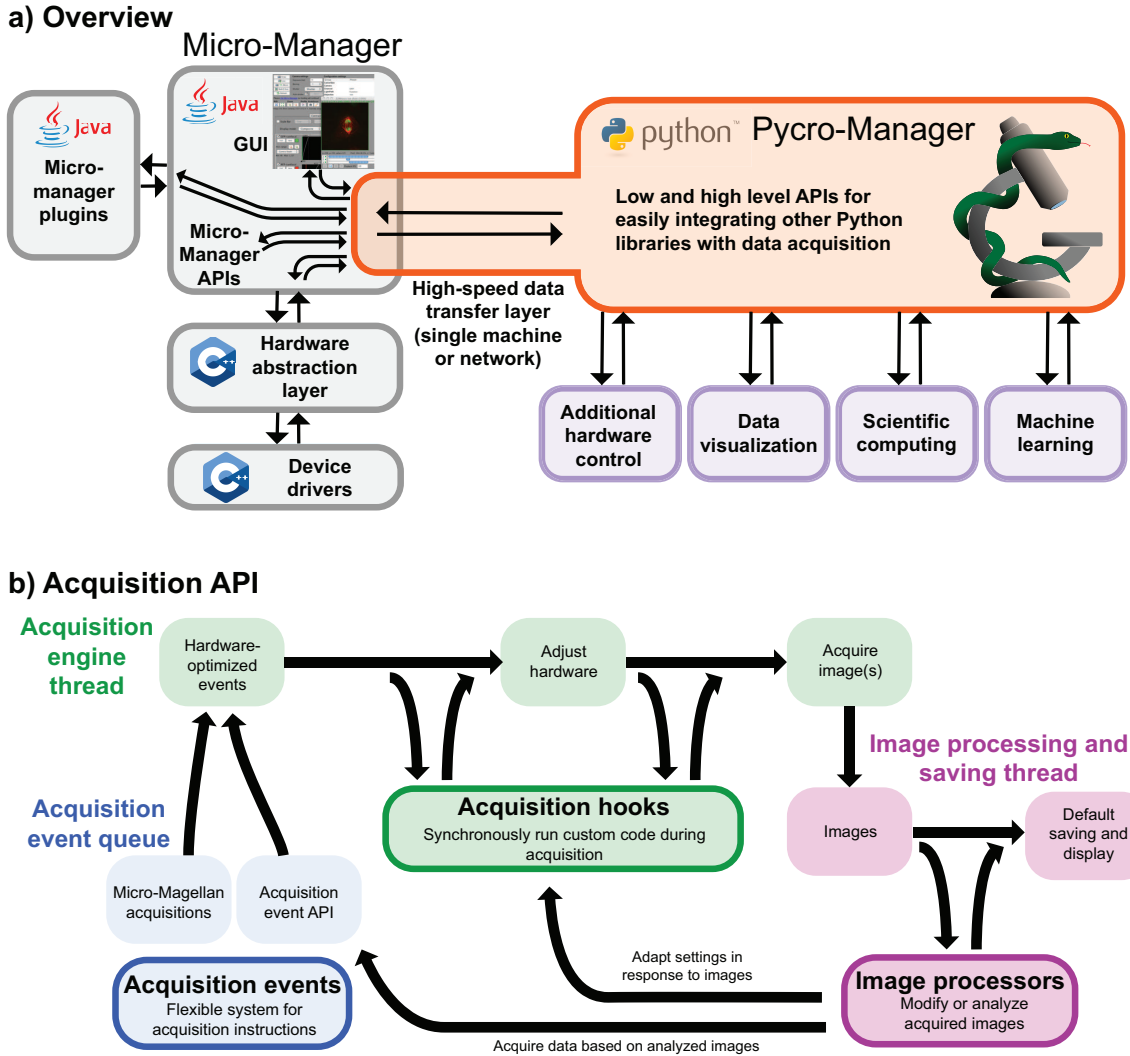}
\caption{\textbf{a) Pycro-manager overview.} The grey boxes denote the C++ and Java components of \(\mu\)Manager, including the GUI, Java APIs, and a hardware abstraction layer that enables generic microscope control code to work on a variety of hardware components. The red box shows Pycro-Manager, which consists of a high speed data transfer layer that can operate within a machine or over a network. This layer enables access to all the existing capabilities of \(\mu\)Manager as if they were written in Python, so that they can inter-operate with Python libraries (purple boxes) for hardware control, data visualization, scientific computing, etc. \textbf{b) Pycro-Manager Acquisition API.} The blue boxes show acquisitions starting with some source of "acquisition events", instructions for image collection and associated hardware changes. Green boxes represent acquisition events that are optimized, then used to move hardware and collect images. "Acquisition hooks" can be used to execute arbitrary code synchronously or modify/delete instructions on-the-fly. Magenta boxes represent acquired images going straight to the default image saving and display, or being diverted through "image processors", which allow for modification of images or diversion to external saving and visualization.}
\label{pycromanager}
\end{figure*}

To address this need, we developed Pycro-Manager. The foundation of Pycro-Manager is a high-speed data transfer layer that dynamically translates between Java and Python (Fig. \ref{pycromanager}a), enabling users to call Java libraries as if they had been written in Python. This creates an easy access point through Python for all the existing capabilities of \(\mu\)Manager, without requiring rewriting of any code or developers in the Java and Python communities to switch languages. The transfer layer can also be run over a network, enabling, for example, splitting out of control and processing to a different machine than the one connected to the microscope.

On top of this layer, Pycro-Manager implements an acquisition API with the flexibility to independently customize what is acquired ("acquisition events"), what happens while it is being acquired ("acquisition hooks"), and what happens to the data after it is acquired ("image processors"), without having to write all the boilerplate code that usually accompanies such customized experiments (Fig. \ref{pycromanager}b). This API can be used, for example, to synchronize external hardware with the acquisition process, modify acquired images on-the-fly before saving/visualization, or implement customized data saving/visualization. Combining these features can be used for even more powerful applications, such as data-adaptive microscopy.

Acquisition events describe what to acquire. For example, a $z$-stack would be a series of acquisition events, each corresponding to a single image and containing the position of the focus drive. They can be created either from a GUI (such as Micro-Magellan \cite{Pinkard2016}) or programmatically through the acquisition event API. 

Acquisition hooks enable the synchronization of arbitrary code with acquisition operations. For example, they can be used to add in control of external devices, or apply auto-focusing operations before proceeding with acquisition.

Image processors give access to image data and metadata as soon it is acquired. They can be used to modify data before reentering the default saving/visualization pipeline, or to divert it to custom alternatives. They can also be used in conjunction with acquisition hooks to create feedback loops for updating hardware based on data, or with acquisition events for controlling future data acquisition.

In summary, Pycro-Manager provides a much needed interface between over a decade's worth of open source microscopy development compatible with the cutting edge of scientific computing. It removes the need to write boilerplate code when developing new types of smart microscopes, freeing developers to rapidly test new ideas. Finally, it does this in a hardware-agnostic way to maximize the portability of code across different microscopes.

The source code for Pycro-Manager can be found at: \url{https://github.com/micro-manager/pycro-manager}, and the documentation is at: \url{https://pycro-manager.readthedocs.io/en/latest/#}

\paragraph*{Acknowledgements}
We thank Stéfan van der Walt for helpful discussion during development and and Kyle Marchuk for early beta testing. 

\bibliography{references}
\bibliographystyle{plain}

\end{document}